\documentclass[twocolumn,a4paper,preprintnumbers,amsmath,amssymb,floatfix,showkeys,showpacs,prl]{revtex4}

\usepackage{graphicx}% Include figure files
\newcommand{\bm}[1]{\mbox{\boldmath$#1$}}

\newcommand{\B}{\mbox{\boldmath $B$}}

\begin{document}

\title{Quantum States of Neutrons in Magnetic Thin Films}

\author{F. Radu$^1$}
\email{florin.radu@rub.de}
\homepage{http://www.ep4.ruhr-uni-bochum.de}
\author{V. Leiner$^{2}$}
\author{M. Wolff$^{1,3}$}
\author{V. K. Ignatovich$^4$}
\author{H. Zabel$^1$}

\affiliation{$^1$Department of Physics, Ruhr-University Bochum, D-
44780 Bochum, Germany} \affiliation{$^2$ Institut f\"ur
Werkstoffforschung WFN, GKSS Forschungszentrum GmbH, 21502
Geesthacht, Germany} \affiliation{$^3$Institut Laue-Langevin,
F-38042 Grenoble Cedex 9, France} \affiliation{$^4$Frank
Laboratory of Neutron Physics, Joint Institute for Nuclear
Research, 141980, Dubna Moscow Region, Russia}

\begin{abstract}

We have studied experimentally and theoretically the interaction
of polarized neutrons with magnetic thin films and magnetic
multilayers.  In particular, we have analyzed the behavior of the
critical edges for total external reflection in both cases. For a
single film we have observed experimentally and theoretically a
simple behavior: the critical edges remain fixed  and the
intensity varies according to the angle between the polarization
axis and the magnetization vector inside the film. For the
multilayer case we find that the critical edges for spin up and
spin down polarized neutrons move towards each other as a function
of the angle between the magnetization vectors in adjacent
ferromagnetic films. Although the results for multilayers and
single thick layers appear to be different, in fact the same
spinor method explains both results. An interpretation of the
critical edges behavior for the multilyers  as a superposition of
ferromagnetic and antifferomagnetic states is given.

\end{abstract}
\pacs{74.78.Fk, 61.12.Ex,  75.25.+z} %\maketitle

\date{\today}
\keywords{Magnetic multilayers, polarized neutron scattering}
\maketitle

%%      Your contribution follows:
%%      DO NOT USE subsection and subsubsection

\section{Introduction}
Neutron reflectivity goes back to 1946 when it was first used by
Fermi and Zinn for the determination of coherent scattering
lengths~\cite{ferm}. Subsequently inserting a polarizer and
analyzer to produce a polarized neutron beam, Hughes and Burgy
started to perform polarized neutron experiments as early as 1951
~\cite{hugh}. It was foreseen by the authors of this work that
\textit{beams of completely polarized neutrons will be useful in
the study of magnetic and nuclear properties}. The next major
achievement of Polarized Neutron Reflectometry (PNR) was the
prediction of spin-flip reflectivity by Ignatovich
(1978)~\cite{ignatovich1978} and the pioneering experiments on
magnetic surfaces %its experimental observation
by Felcher~(1981)~\cite{felcher1981}. While specular PNR is widely
recognized as a powerful tool for the investigation of
magnetization profiles in magnetic heterostructures
\cite{Fitzsimmons}, the description of off-specular scattering
from magnetic domains is still under development
\cite{suzane2001}. In spite of these important developments there
is still a confusion concerning the quantum states of neutrons in
a magnetic sample. Here we show unambiguously that the neutron has
to be treated as a spin 1/2 particle~\cite{ples,radu} in each
homogeneous magnetic layer. This is at variance with the
conventional description of neutron reflectivity, which often
considers the neutron magnetic potential as a classical dot
product~\cite{Felcher,majk,zabel94,Temst}.

Neutrons interact with a magnetic thin film via the Fermi nuclear
potential and via the magnetic induction. Thus, the neutron - film
interaction hamiltonian includes both contributions:
$V=V_n+V_m=(\hbar^2/2m) 4\pi N b-{\bm{\mu B}}$, where $m$ is the
neutron mass, $N$ is the particle density of the material, $b$ is
the coherent scattering length, $|{\bm{\mu}}|$ is the magnetic
moment of the neutron, and $|{\bm{B}}|$ is the magnetic induction
of the film. Unconventionally, however, neutron reflectivity
treats the dot product between the magnetic induction and neutron
magnetic moment classically: $V_m=-\bm{\mu B}=\pm \bm{|\mu|
|B|}\cos(\theta)$, where $\theta$ is the angle between the
incoming neutron polarization direction and  the direction of the
magnetization inside the film. Writing the magnetic potential as a
classical dot product implies that the neutron energies in the
magnetic layer have a continuous distribution from -$|\mu| |B|$ to
+~$|\mu||B|$. This predicts that the critical angle for total
reflection depends on the angle between the direction of
polarization and the direction of the magnetic field inside the
layer:
\begin{equation}
\frac{4\pi
\sin(\alpha_c^{\pm})}{\lambda}=Q^{\pm}_c=\sqrt{\frac{2m}{\hbar^2}
\,  (V_n\pm
 \bm{|\mu|| B_s|} \cos(\theta)) }, \label{eq1}
\end{equation}
where $\alpha$ is the glancing angle to the surface, $\lambda$ is
the wavelength of the neutrons, and  Q$^{\pm}_c$ is the critical
scattering vector. There  are experimental data \cite{schr,majk}
on magnetic multilayers  which apparently confirm this behavior.
Therefore, the classical representation appears to provide a
convenient and transparent way to describe the experimental
observations ~\cite{Felcher,majk,zabel94,Temst}.

From the Stern-Gerlach experiment we know that there are only two
eigen states for the spin 1/2 particles in a magnetic field.
Therefore, the eigen wave number of a neutron in a magnetic thin
film has two proper values. After solving the Schrodinger equation
one obtains two eigen wave numbers for neutrons in a magnetic
film: $k^2_{\pm}=\frac{2m}{\hbar^2}(V_n \pm \bm{|\mu| |B|})$. They
correspond to two possible states of spin orientation: one for the
case, when the spin is parallel to the magnetic induction, and the
other one for the antiparallel orientation. %Note that the
%quantization axis $\bm{Z}$ is always parallel to the magnetic
%induction \textbf{B}.
 It follows that there are only two possible
energies and consequently only two values for the index of
refraction corresponding to the spin-up and spin-down states of
the neutrons. Therefore, QM predicts that there are only two
critical angles for the total reflection: one corresponding to the
R$^+$ and one to the R$^-$ reflectivity
\begin{equation}
\frac{4\pi
\sin(\alpha_c^{\pm})}{\lambda}=Q^{\pm}_c=\sqrt{\frac{2m}{\hbar^2}\,(V_n\pm
 \bm{|\mu|| B_s|)}  } \label{eq2}
\end{equation}
Obviously there is a contradiction between the quantum mechanical
prediction (Eq.\ref{eq2}) and the prediction based on the
classical representation of the magnetic potential(Eq.\ref{eq1}):
quantum mechanics predicts that the spin states of the neutron is
determined by the magnetic induction in the sample, whereas
classical representation of the magnetic potential, supported by
experiments on magnetic multilayers, assert that the spin states
of the neutrons is fixed by the incident polarization axis.

%Obviously there is a contradiction between the quantum mechanical
%prediction and the experimental observation: quantum mechanics
%predicts that the spin state of the neutron is determined by the
%magnetic induction in the sample, whereas experiments on magnetic
%films and multilayers assert that the spin state of the neutrons
%is fixed by the incident polarization axis.

Here we describe an experiment which provides direct and
unambiguous evidence for the spin states of neutrons in magnetic
media. The goal is to find a system where the angle between the
neutron polarization and direction of the magnetization inside of
the film can be fixed and controlled. Then we measure the R$^+$
and R$^-$ reflectivities and determine whether the position of the
critical edges changes as a function of the angle $\theta$, or
whether the critical edges stay fixed, and only intensity
redistributes between reflections R$^+$ and R$^-$ with change of
$\theta$. The easiest way to control the angle $\theta$ is to
rotate the magnetic film and therefore the magnetization direction
with respect to the neutron spin polarization, which remains fixed
in space outside of the sample. This requires that the film should
have a high remanent magnetization. Additionally, the film
thickness should exceed the average neutron penetration
depth~\cite{bally}. The last requirement is essential in order to
avoid neutron tunnelling effects which will hinder the precise
determination of the critical edges.

\section{Sample characterization by MOKE and PNR}

\begin{figure}[!h]
\begin{center}
\includegraphics[clip=true,keepaspectratio=true,width=1\linewidth]{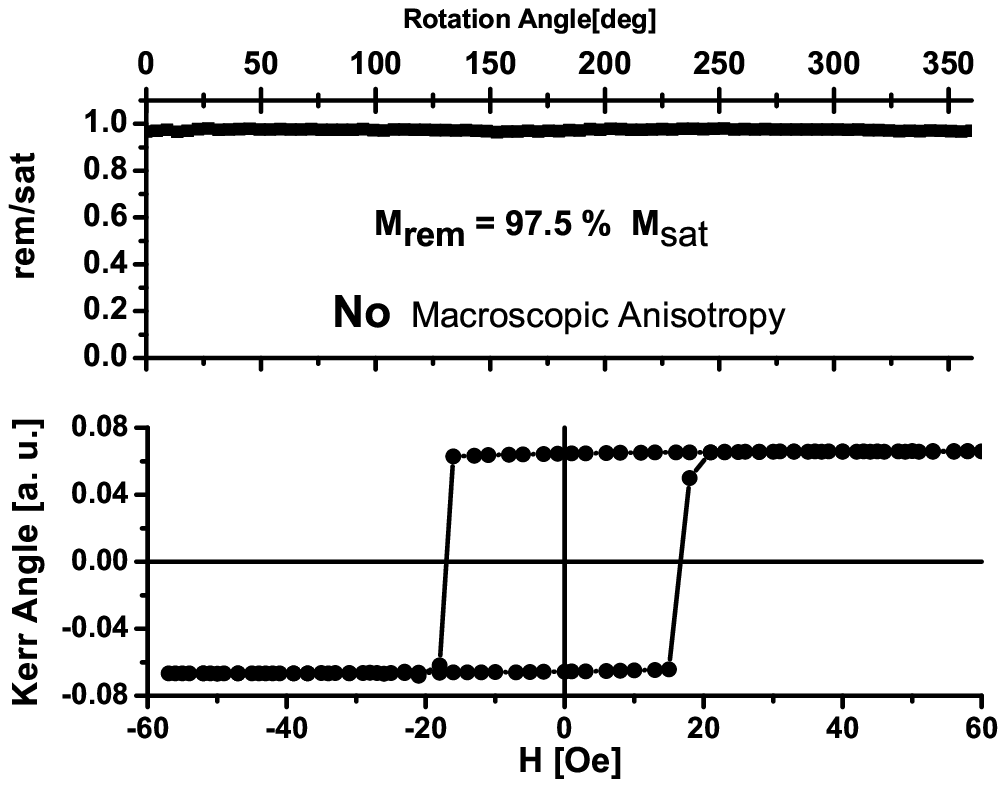}
\end{center}
\vspace{-.5cm} \caption{\label{mokeFe}  Bottom: hysteresis loop of
the polycrystalline Fe/Si sample measured by MOKE. Top: the
behavior of the remanent magnetization as a function of the
rotation angle extracted from hysteresis loops. }
%\end{figure}

%\begin{figure}[!h]
\begin{center}
\includegraphics[clip=true,keepaspectratio=true,width=.8\linewidth]{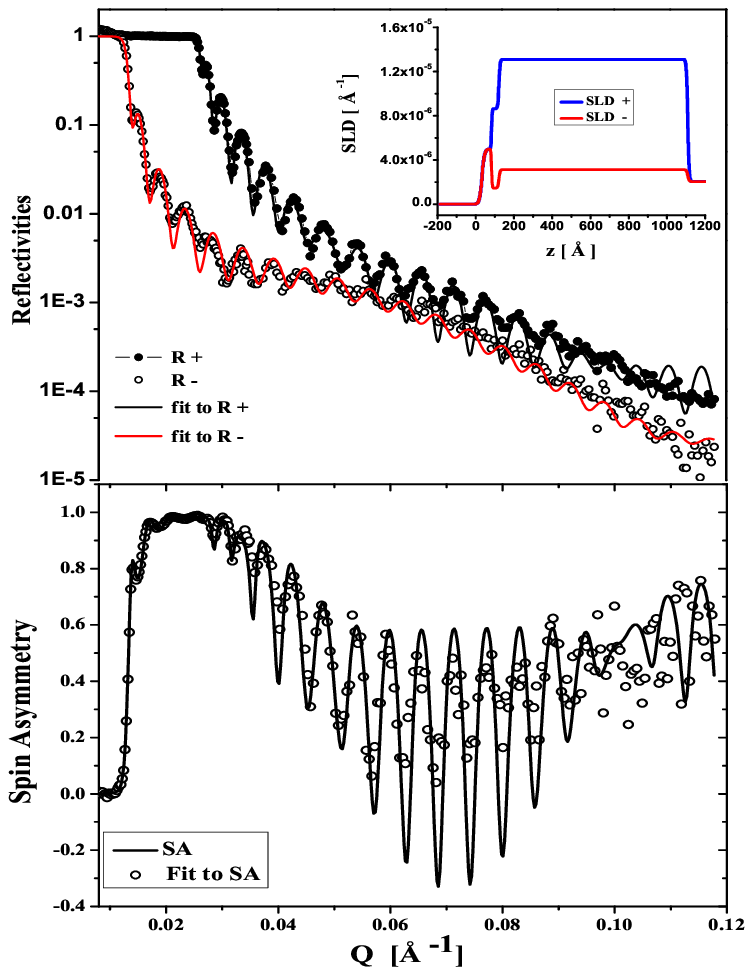}
\end{center}
\caption{\label{pnr1} (Color online). Top: Polarized neutron
reflectivity curves R$^+$ (solid black circles) and R$^-$ (open
black circles) of the Fe/Si sample. The black line is the
simulated R$^+$ reflectivity and the red (light gray) line is the
simulated R$^-$ reflectivity. The applied magnetic field was
2000~Oe. In the inset, the magnetic profile obtained from fitting
the data is shown. Bottom: Experimental (open black symbols) and
simulated (black line) spin asymmetry ($(R^+ \, - \, R^-)/(R^+ \,
+ \, R^-)$) are plotted for the same sample in saturation. All
lines in the top and bottom panels are fits to the data points
using the GMM (for more details see text). The abscissa is the
wave vector transfer: $Q= 4 \pi \sin(\alpha)/\lambda$. }
\end{figure}

To fulfill the aforementioned requirements, we have chosen a 100
nm thick polycrystalline Fe film deposited by rf-sputtering on a
Si substrate. The thickness of the Fe films was about 4 times
larger than the average penetration depth
$1/\sqrt{2mV_N/\hbar^2}$.  The Fe film was covered with thin Co
and CoO layers, the latter one protecting the Fe film from
oxidation. For sample characterization at room temperature we
first recorded hysteresis loops with the magneto-optical Kerr
effect (MOKE). A series of hysteresis loops were taken with the
field parallel to the film plane but with different azimuth angles
of the sample. A typical hysteresis loop is shown in
Fig.~\ref{mokeFe}.  The coercive field is about 20~Oe and the
remanence is high. A plot of the ratio between the remanent
magnetization and saturation magnetization $M_{rem}/M_{sat}$
versus the rotation angle about the sample normal is shown in
Fig.~\ref{mokeFe}. We conclude that the system has no macroscopic
anisotropy and the remanent magnetization is 97.5\% of the
saturation magnetization.

Neutron reflectivity experiments were performed using the angle
dispersive neutron reflectometer ADAM installed at the Institut
Laue-Langevin, Grenoble, which operates at a fixed wavelength of
4.41 \AA. The R$^+$ and R$^-$ reflectivities and the spin
asymmetry (R$^+$ - R$^-$)/(R$^+$ + R$^-$) in a saturation field of
2000~Oe are plotted in the top and bottom panel, respectively, of
Fig.~\ref{pnr1}. The solid lines are fits to the data points using
the {\it{PolarFit}} code based on the general matrix method (GMM)
~\cite{radu}. The fit and sample parameters are listed in
Table~\ref{table1}. In order to obtain a high confidence of the
fit parameters, all reflectivities were fittted together and with
the same parameter set. In general it is useful to fit first the
spin asymmetry, for which geometrical and normalization parameters
drop out.

 \begin{table}[!ht]
 \caption{\label{table1} Parameters of the Fe/Si sample obtained by fitting
to the R$^+$ an R$^-$ data shown in Fig.\ref{pnr1}.
$d$ is the layer thickness, $\sigma$ is the rms roughness, {\it SLD} is
the scattering length density, and B is the magnetic induction in the ferromagnetic
films. }
\begin{center}
\begin{tabular}{|l|l|l|l|l|}
\hline

Layer & $d$ [\AA] & $\sigma$[\AA] &$SLD$ & $B [Oe]$ \\
\hline\hline

Co$_y$O$_{1-y}$ & 50 & 11.5 & 4.99e-6 & 0  \\\hline

Co$_x$ Fe$_{1-x}$ & 39 & 3 & 5.0518e-6 & 15563.4  \\\hline

Fe & 987 & 5 & 8.024e-06 & 21600  \\\hline

substrate & non & 6 & 2.073e-06 & 0  \\ \hline
\end{tabular}
\end{center}
\end{table}

From Table~\ref{table1} and the magnetic characterization
(Fig.~\ref{mokeFe}) we conclude that the sample fulfills the
requirements as concerns thickness, anisotropy, and remanence as
required for our experiment.

\onecolumngrid

\begin{figure}[!h]
    \begin{center}
      \includegraphics[clip=true,keepaspectratio=true,width=.8 \linewidth]{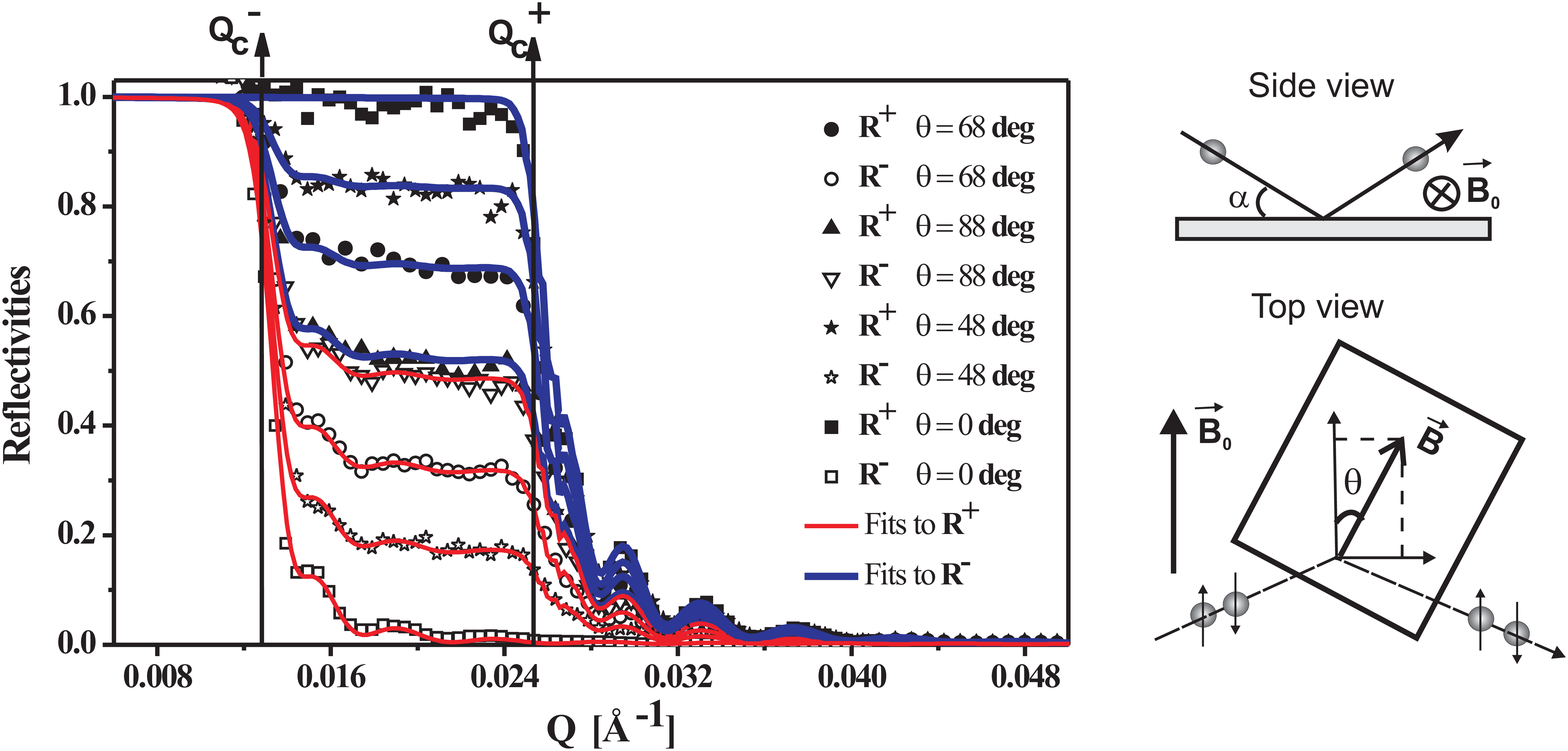}
      \caption{(Color online). Experimental (symbols) and
      simulated (lines)
       reflectivity curves (R$^+$ and R$^-$) from Fe(1000 \AA)/Si
       sample. The abscissa is the wave-vector transfer. The two
       sets of $R^+$ (solid  black symbols) and $R^-$ (open black symbols)
       reflectivity curves  were measured
       for four different $\theta$ angles between the neutron
       polarization along $\bm{B_0}$
       and the direction of the
         magnetic induction ($\bm{B}$) which lies in the sample plane.
         The guiding field is  $B_0\approx 10$ Oe. The blue  (thick dark gray) lines
         are
the simulated R+ reflectivities and the red (thin light gray)
lines are the simulated R- reflectivities. In the right side the
        experimental geometry is shown.
                The figure shows that the critical edges $Q^c_{+}$ and $Q^c_{-}$ are not sensitive to the $\theta$ angle.}
      \label{PNRrot1}
    \end{center}
\end{figure}

 \twocolumngrid

\section{Rotation experiment}

 The rotation experiment was performed as follows: the Fe layer was
magnetized parallel to the neutron polarization direction and then
the magnet was removed. A small guiding field (H$_c$) is still
present at the sample position in order to maintain the neutron
polarization. Subsequently a series of R$^+$ and R$^-$
reflectivities, shown in Fig.~\ref{PNRrot1}, were measured for
several in-plane rotation angles of the sample. We observe two
characteristics of the reflectivities: (1) the critical edges are
fixed and independent of the in-plane rotation angle $\theta$ (see
Eq.~\ref{eq2}), and (2) the R$^-$ intensity continuously increases
at the expense of the R$^+$ intensity as a function of the
$\theta$ angle. Using the parameters obtained from the fit  to the
saturation data (see Table~\ref{table1}) and using the rotation
angle $\theta$ as set during the experiment, we have simulated
~\cite{polar} the reflectivities using the GMM
approach~\cite{radu} which transparently predicts the behavior of
the critical edges described by Eq.~\ref{eq2}. The simulated
curves are plotted together with the experimental data in
Fig.~\ref{PNRrot1}. There are no free parameters for these
simulations, providing an excellent description of the
experimental results. The fixed critical edges $Q^c_{+}$ and
$Q^c_{-}$ can easily be interpreted in the context of the neutron
spin states in homogenous magnetic media as discussed in the
introduction.

\section{Experimental determination of magnetization orientation}

%During the spin-flip process the neutron will change its spin
%direction from parallel alignment with the magnetic induction to
%antiparallel, or vice versa. During this process it can gain or
%loose the Zeeman energy independent of the rotation angle. The
%influence of the angle on the R$^+$ and R$^-$ can be described as:
%$nSA(\theta)=\frac{SA(\theta)}{SA(0)}=cos(\theta)$, where
%$nSA(\theta)$ is the normalized spin asymmetry~\cite{prb2003}.

\begin{figure}[!ht]
    \begin{center}
\includegraphics[clip=true,keepaspectratio=true,width=1 \linewidth]{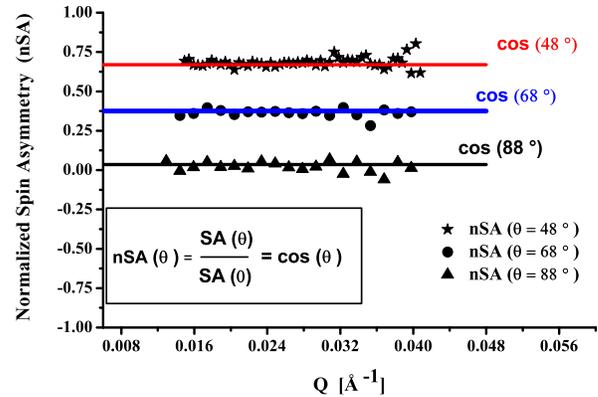}
      \caption{(Color online). Solid black symbols: The experimental normalized spin asymmetries ($nSA (\theta)=SA(\theta)/(SA (0))$),
      plotted as a function of the wave-vector transfer Q.
        Lines:
      The three  lines are the cosines of the corresponding angles set
      during the experiment.
      From top: cosine of 48$^\circ$ (thin gray (red) line) , 68$^\circ$ (thick gray (blue) line), and 88$^\circ$ (thin black line),
      respectively. The angles are the experimental rotation angles $\theta$ used during the experiment shown in Fig.~\ref{PNRrot1}.
      The experimental normalized spin asymmetries are assembled from the $R^+$ and
      $R^-$ reflectivities shown in Fig~\ref{PNRrot1}. This figure shows that the equation $nSA(\theta$)=cos($\theta$)
      is valid over the whole Q range for a single magnetic layer.
      It can be used to extract the angle $\theta$ directly from the experimental reflectivities.
 }
      \label{PNR_SA1}
    \end{center}
\end{figure}

The sensitivity to the in-plane rotation angle of the
magnetization is seen very clearly seen in the reflected
intensities R$^+$ and R$^-$ plotted in Fig.~\ref{PNRrot1} . It has
been shown theoretically~\cite{prb2003} that, for a single
magnetic layer, the normalized spin asymmetry ($nSA(\theta)$) is
directly related to the $\theta$ angle through the following
expression:
\begin{equation}
nSA(\theta)=\frac{SA(\theta)}{SA(0)}=\cos(\theta) \label{SAeq}
\end{equation}
 Now, we use our
experimental data shown in Fig.~\ref{PNRrot1} to confirm the
validity of this equation. In Fig.~\ref{PNR_SA1} is shown the
experimental normalized spin asymmetry and the cosine of the
experimental angles. The agreement between the experimental
normalized spin asymmetry (symbols) and the cosine of the $\theta$
angles (lines) set during the experiment is excellent over the
whole wave vector transfer range.  It follows that the
magnetization orientation of a single magnetic layer with respect
to the neutron polarization outside the layer can be easily
extracted experimentally  using Eq.~\ref{SAeq}. For more
complicated systems a numerical fitting is still necessary. The
$nSA$ is an important measure of  hysteresis loops. It was shown
in Ref. ~\cite{prb2003} that $nSA$ can be written, generally, as:
$nSA=M_{||}/M_{sat}$ for both, magnetization reversal via coherent
rotation and via domain wall movement. This implies that $nSA$
reproduces the hysteresis loops as measured by SQUID or MOKE. Here
we confirm the validity of the $nSA$ for determining the
magnetization reversal via coherent rotation. We mention that this
equation is valid for samples which contain a single magnetic
layer. Comparing MOKE or SQUID hysteresis loops with $nSA$ is a
very useful tool for the evaluation of magnetic domain state
and/or a reduced magnetization within the layer.

\section{Multilayers}

\begin{figure}[!h]
\begin{center}
\includegraphics[clip=true,keepaspectratio=true,width=1\linewidth]{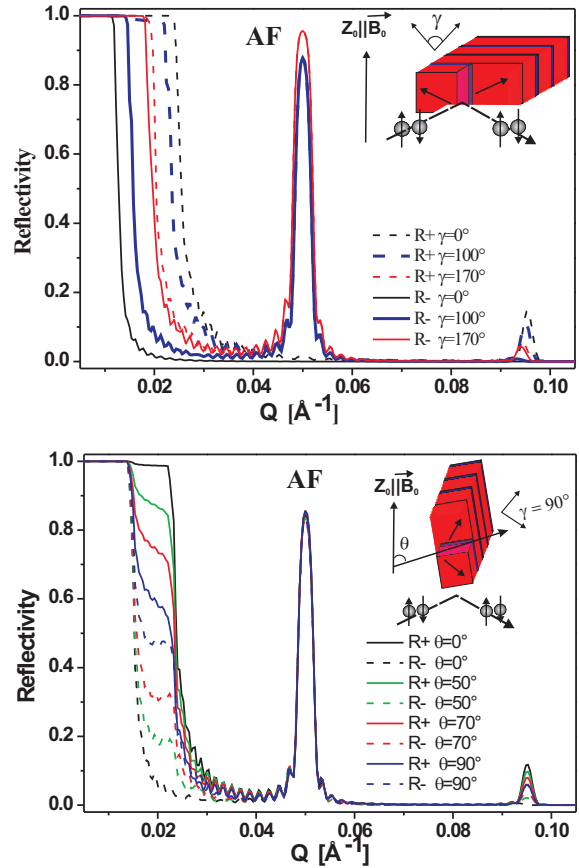}
\end{center}
\vspace{-.5cm} \caption{\label{FeCr} (Color online). Top:
Simulation of polarized neutron reflectivities [R$^+$ (dashed
lines, shifting from right to left) , R$^-$ (solid lines, shifting
from left to right)] for a Fe/Cr multilayer as a function of
coupling angle ($\gamma$) between the magnetization vectors of
adjacent Fe layers. Bottom: Simulations of R$^+$ (solid lines,
shifting from top  down) and R$^-$ (dashed lines, shifting from
bottom up) as a function of rotation angle $\theta$ for
$\gamma=90^\circ$ }
\end{figure}

Our next topic is  to investigate the neutron spin states in
multilayers with noncollinear magnetization of adjacent layers. We
simulated the reflectivity profile of a
[Fe(60\AA)/Cr(8\AA))]$_{40}$/Si superlattice, with thicknesses of
the Fe and Cr layers which are typical for many real
superlattices~\cite{schr}. For the simulation we used the freeware
code {\it{PolarSim}}~\cite{polar} based on GMM for the calculation
of the reflection and transmission coefficient together with a
full quantum mechanical description of the spin
states~\cite{radu}. In the simulation the choice of a Si substrate
has the advantage that it does not obscure the critical edge of
the (-) neutron state. In the top panel of Fig.~\ref{FeCr} we show
simulations of R$^+$ and R$^-$ reflectivities for three angles
$\gamma$ between the magnetization vectors of adjacent Fe films:
$\gamma=0$ (or ferromagnetic alignment); $\gamma=100^\circ$; and
$\gamma=170^\circ$ (close to antiferromagnetic alignment). Our
focus is on the behavior of the critical scattering vector for
total reflection. We observe that for $\gamma=0$ the (+) and (-)
critical scattering vectors are well separated and that they
contain information about the saturation magnetization. When the
$\gamma$ value increases, the critical edges approach each other.
For an angle $\gamma=180^\circ$ (not shown here) there is no
difference between the R+ and R- reflectivities. The main result
from this simulation is the observation that the separation of the
critical edges is a continuous function of the angle $\gamma$
between the in-plane adjacent magnetization vectors. The critical
edge positions satisfy the following relation:
\begin{equation}
\frac{4\pi
\sin(\alpha_c^{\pm})}{\lambda}=Q^{\pm}_c=\sqrt{\frac{2m}{\hbar^2}\,
( V_n^{eff}\pm
 \bm{|\mu|| B_s|} \cos(\gamma/2)) }, \label{eq3}
\end{equation}
where $V_n^{eff}$ is an effective nuclear potential. Clearly, for
this geometry the angles $\theta$ and $\gamma/2$ coincide if the
neutron polarization is parallel to the average field
$(\B_1+\B_2)/2$, where $B_1$ and $B_2$ are the magnetic field
inductions of adjacent layers. Therefore, numerically, the eqs. 1
and 3 are almost identical. However, there is a fundamental
difference: similarly to the single layer, the $\theta$ angle does
not influence the position of critical edges, whereas the $\gamma$
angle is solely responsible for the continuous shift.

To shed more light on how the $\theta$ and $\gamma$ angles affect
the critical edges for polarized neutron reflectivity at the
multilayers we simulated numerically the rotation experiment
performed on the single layer. For a fixed coupling angle of
$\gamma=90°$, as it can be achieved also experimentally via
biquadratic exchange coupling, the reflectivities R$^+$ and R$^-$
are plotted as a function of the $\theta$ angle.  Here the
$\theta$ angle is the angle between the incoming neutron
polarization and the direction of the average magnetization vector
of two adjacent ferromagnetic layers. The results are shown in the
bottom panel of Fig.~\ref{FeCr}. We observe a similar behavior of
the critical edges and intensities as for the single layer. While
the positions of the critical scattering vectors $Q^c_{+}$ and
$Q^c_{-}$ remain fixed for a constant coupling angle $\gamma$, the
R$^-$ intensity increases on the expense of the R$^+$ intensity
with increasing $\theta$ angle. With this simulations we lift the
contradiction stated in the introduction by showing that
Eq.~\ref{eq1} is a particular case of Eq.~\ref{eq3}, which, in
turn, is in agreement with the QM description of the neutron spin
states in magnetic media Eq.~\ref{eq2}. The  different behaviors
of the critical edges for the case of a single homogeneous
ferromagnetic layer and for a multilayer with alternating
directions of the layer magnetization vectors now becomes obvious:
in the multilayer the neutrons are affected by an average magnetic
potential which depends on the relative orientation of the
magnetic induction in the individual layers. However, in both
cases, single film as well as multilayer, the magnetic potential
of the individual layers ($V_m=|\mu| |\bm{B_s}|$) enters the
algorithm for calculating the reflectivities.

It should be noted that the dependence of $Q^c_{+}$ and $Q^c_{-}$
on the angle $\gamma$ in a multilayer  is a general property of
the periodic potential with different field orientation and
magnitude. It is natural to expect that such a sample is a
noncollinear ferrimagnet with ferromagnetic field
$\B_f=(\B_1+\B_2)/2$,
\begin{equation}\label{ma5}
B_f=|\B_1+\B_2|/2=B\cos(\gamma/2),
\end{equation}
and with antiferromagnetic field $B_{af}=(\B_1-\B_2)/2$,
\begin{equation}\label{ma6}
B_{af}=|\B_1-\B_2|/2=B\sin(\gamma/2).
\end{equation}
Then, the critical edges can be expected to be given by
Eq.~\ref{eq3}.

To further stress the origin of the  effective nuclear potential
$V_n^{eff}$ term in Eq.~\ref{eq3}., let us consider the critical
edge for non-polarized neutrons when scattered at a [Fe(x
\AA)/Cr(y \AA))]$_{\infty}$ multilayer. Naively, we may expect
that the critical edge to be given by the Fermi interaction
potential of Fe as it is higher than the
 potential of Cr. This is, however, not the case. For  a finite
 thickness x of the Fe layer and zero thick Cr layer, indeed the critical edge
 is equal to the critical edge of a single thick Fe layer.
 Vice versa, for  zero
 thickness of Fe layer and finite thickness y for the Cr layer the critical edge is
 given by the Fermi potential of Cr. However, when both layers have
 finite thicknesses the critical edge of the multilayer will vary
from the value for pure Fe to the value for pure Cr. Therefore,
the  critical edge for non-polarized neutrons reflected from a
multilayer not only depends on the Fermi potential of the two
separate layers, but also  on their individual thicknesses. %In
%Eq.~\ref{eq3} the term $V_n^{eff}$
% accounts for such an effective Fermi potential which provides the
%  critical edge for non-polarized neutrons.

\section{Conclusions}
In summary, we have analyzed the behavior of the critical
scattering vectors $Q^c_{+}$ and $Q^c_{-}$   for total external
reflection of a polarized neutron beam for the case of homogeneous
ferromagnetic films and for antiferromagnetically coupled
multilayers. For a single film we have observed experimentally and
shown theoretically that the critical edges do not change as a
function of the angle between the neutron polarization and the
direction of the magnetic spins inside the film. They fulfill the
relation Eq.~\ref{eq2}: $Q^{\pm}_c=\sqrt{\frac{2m}{\hbar^2}\, (
V_n\pm
 \bm{|\mu|| B_s|)}}$, which directly reflects the
spin states of the neutron beam in magnetic thin films. For
multilayers we found that the critical edges for total external
reflection move towards each other as a function of the coupling
angle. Their position is well reproduced by the Eq.~\ref{eq3}:
$Q^{\pm}_c=\sqrt{\frac{2m}{\hbar^2}\, (V_n^{eff}\pm  \bm{|\mu||
B_s|} \cos(\gamma/2)) }$. The $\cos(\gamma/2)$ dependence is not
related to the neutron spin states in the magnetic media, but it
is the result of the presence of a ferromagnetic field direction
along the average field in the noncollinear ferrimagnetic. By
choosing a fixed coupling angle $\gamma$ between the magnetization
vectors of adjacent layers and rotating the sample, the critical
edges behave  again in accordance with the neutron spin states in
homogeneous magnetic media. Practically, the coupling angle in
non-collinear superlattices can be inferred directly from the
experimental data through the separation of the critical edges.
For a single layer the orientation of the magnetization can be
extracted experimentally from the spin asymmetry.

%Most importantly, we have shown that in either
%case, film and multilayer, the magnetic potential cannot be treated vectorially.

This work was supported by the Deutsche Forschungsgemeinschaft
through the research network (SFB 491): \textit{Magnetic
heterostructures}, which is gratefully acknowledged. We would like
to thank Sabine Erdt-B\"{o}hm for the sample preparation. The ADAM
reflectometer is supported by BMBF contract O3ZA6BC1.

\newpage

\end{document}